\documentclass[aps,superscriptaddress,eqsecnum,nofootinbib,showpacs,preprintnumbers]{revtex4-2}
\usepackage[utf8]{inputenc}
\usepackage{float}
\usepackage{hyperref}
\usepackage{amsmath}
\usepackage{amsfonts}
\usepackage{latexsym}
\usepackage{xcolor}
\usepackage{graphicx}
\usepackage{float}
\usepackage{xcolor}
\usepackage{framed}
\usepackage{amssymb}
\usepackage{graphicx, rotating}
\usepackage{epsfig}
\usepackage{latexsym}
\usepackage{amsmath,bm,amssymb}
\usepackage{slashed}
\usepackage{hyperref}
\usepackage{epstopdf}
\usepackage{slashed}
\usepackage{tabularx}
\begin{document}
\title{Scalarization of the Reissner-Nordstr\"om black hole with higher derivative gauge field corrections
}

	\author{Stella Kiorpelidi}
	\email{<stellakiorp@windowslive.com>}
	\affiliation{Physics Department, National Technical University of Athens, 15780 Zografou Campus, Athens, Greece}

	\author{Thanasis Karakasis}
	\email{<thanasiskarakasis@mail.ntua.gr>}
	\affiliation{Physics Department, National Technical University of Athens, 15780 Zografou Campus, Athens, Greece}

	\author{George Koutsoumbas}
	\email{kutsubas@central.ntua.gr}
	\affiliation{Physics Department, National Technical University of Athens, 15780 Zografou Campus, Athens, Greece}
	
	\author{Eleftherios Papantonopoulos}
	\email{lpapa@central.ntua.gr} \affiliation{Physics Department, National Technical University of Athens, 15780 Zografou Campus, Athens, Greece}
	
\begin{abstract}
    We discuss spontaneous scalarization of the Reissner-Nordstr\"om black hole in the presence of higher derivative gauge field corrections that arise in the context of string, as well as higher-dimensional more fundamental gravity theories. Our theory admits the Reissner-Nordstr\"om solution at the scalar vacuum of the theory ($\phi=0$) and we find that the higher order derivative gauge field correction term results in the tachyonic instability of our system once the coupling function satisfies the condition that its second derivative is positive at the scalar vacuum in the appropriate parameter space. We find that the branches do not end with an extremal black hole, rather with a singularity as indicated by the divergence of the Kretschmann scalar. The black holes can be overcharged in the sense that they may carry larger electric charge in comparison to their mass. Finally, these solutions possess larger entropy at the event horizon radius when compared to the Reissner-Nordstr\"om black hole, as well as to scalarized black holes without the higher order derivative gauge field terms, indicating in this way the thermodynamic prefer-ability of our system, when compared to existing literature, while they respect the energy conditions.
\end{abstract}
\maketitle
\tableofcontents

\section{Introduction}

The process of spontaneous scalarization has been proposed by Damour and Esposito-Far\'ese in the context of neutron stars \cite{Damour:1996ke}. Their investigation showed that a particular coupling between gravity and a scalar field, leads to physical phenomena that are indistinguishable from the results of General Relativity (GR) when gravity is weak, but predicts strong deviations from the results predicted by GR in the strong field regime of neutron stars. Their analysis is consistent with the observations that one can make in the weak field limit, where it seems that there is no other fundamental field at play besides the metric tensor. However, because of the technological advancement and routine detection of gravitational waves by the LIGO-Virgo-Karga Collaboration \cite{LIGO}, one is now enabled to study the existence of gravity-related fundamental fields besides the metric tensor, that seem to be elusive in the weak field limit. For such fundamental fields to have remained undetected so far, there has to exist a mechanism that suppresses them in the weak field regime. That mechanism could be described by spontaneous scalarization, which is a mechanism that endows black holes or neutron stars (in general strongly self-gravitating bodies, however in this article we are interested in the black hole scalarization) with a non-trivial scalar field configuration (one can also generalize the mechanism of scalarization to include vector fields (vectorization) or tensor fields (tensorization)), which only appears when a certain quantity that characterizes the self-gravitating body, goes beyond a particular threshold \cite{Doneva:2022ewd}.

The prototype example of black hole scalarization might be found in \cite{Kanti:1995vq}.
Omitting the fourth derivative terms for the dilaton in the Lagrangian that arises in the effective field theory that emanates from the bosonic string, as well as antisymmetric tensor fields, the authors considered the Lagrangian
\begin{equation}
    \mathcal{L} = \frac{R}{2} - \frac{1}{4}(\partial_\mu \phi)^2 + \frac{\mathfrak{\alpha}'}{8g^2}e^\phi\left(R_{\mu\nu\rho\sigma}R^{\mu\nu\rho\sigma} - 4R_{\mu\nu}R^{\mu\nu}+R^2\right)~,
\end{equation}
where $\mathcal{\alpha}'$ is the string tension and $g$ is the string coupling. Here the combination $R_{\mu\nu\rho\sigma}R^{\mu\nu\rho\sigma} - 4R_{\mu\nu}R^{\
    mu\nu}+R^2 \equiv R_{GB}^2$
is the Gauss-Bonnet topological invariant, which would not affect the equations of motion in four dimensions, if not coupled to the dilaton. In \cite{Kanti:1995vq} the authors managed to find hairy black hole solutions, in which the hair is of secondary type, since, the dilaton field does not contribute to the spacetime metric by an additional charge that is independent of the black hole mass (for an example of a primary hair, we refer the reader to \cite{Bakopoulos:2023fmv}). It was then later found that this model might lead to spontaneous scalarization of the Schwarzchild black hole \cite{spsc}, if one considers coupling functions of the scalar field with the Gauss-Bonnet invariant which are quadratic when linearized around the scalar vacuum with respect to $\phi$. The basic idea is that the Schwarzchild black hole solution does solve the theory for a vanishing scalar field $\phi=0$. However, by examining the scalar perturbations in the background of the Schwarzchild black hole it is found that there exists a threshold value for the mass of the black hole, below which the Schwarzchild black hole is no longer stable against scalar perturbation and the spacetime is getting spontaneously dressed by the scalar field of the theory. The key difference between spontaneously scalarized black hole solutions and hairy black holes that can be sourced by a scalar potential for example \cite{scalarhair} is that the dressing with the scalar field in the spontaneous scalarization mechanism is a dynamical procedure while in the hairy black holes generated by a scalar potential \cite{scalarhair}, the potential is engineered in a way to give the Schwarzchild black hole solution (when $\phi=0$) as well as a hairy black hole with a non trivial scalar field, in which the dressing is not dynamical and it is not necessary for the Schwarzchild black hole to be unstable against the scalar perturbations in order for the dressing to happen.

Since these first results, a massive number of articles has been published investigating or generalizing the mechanism of scalarization. While most of them could be found in the review \cite{Doneva:2022ewd}, we will devote this paragraph to some important works.
In \cite{Antoniou:2017hxj} many different forms for the coupling function of the scalar field with the Gauss-Bonnet invariant were considered and the results showed that regular, asymptotically flat black hole solutions may be obtained, that elude the no-scalar-hair theorems. It was also pointed out the scalarized black holes with an exponential coupling function (the dilatonic coupling) will lead to higher entropy when compared to the Schwarzchild black hole. Furthermore, self-interactions have been considered for the scalar field in \cite{Macedo:2019sem,Doneva:2019vuh,Bakopoulos:2020dfg}, with \cite{Macedo:2019sem} providing evidence that a quartic potential for the scalar field can lead to stable scalarized black hole solutions. The pure strong-field regime was considered in \cite{Bakopoulos:2019tvc} and it was found that the pure Gauss-Bonnet term cannot support the existence of black hole solutions. Black hole with a cosmological constant have been considered in \cite{Bakopoulos:2018nui}. The spontaneous scalarization of generalised scalar-tensor theories has also been addressed in \cite{Andreou:2019ikc} as well as the scalarization of a Ricci scalar coupling besides the Gauss Bonnet term \cite{Antoniou:2021zoy}. Spin-induced scalarization is considered in \cite{Herdeiro:2020wei, Dima:2020yac}, while the issue of stability of these types of black holes was also investigated in \cite{stability}. Finally, the effects of mass and self-interaction on non-linear scalarization was tackled in \cite{Pombo:2023lxg}.

The spontaneous scalarization of the Reissner-Nordstr\"om (RN) black hole was explored in detail in \cite{Doneva:2018rou} for different coupling functions between the Gauss-Bonnet invariant and the scalar field and it has been found that the solutions do not reach an extremal limit. The scalarization of the RN black hole with the help of a coupling between the scalar field and the electromagnetic invariant was scrutinized in \cite{Herdeiro:2018wub} in the Einstein-Maxwell-Scalar (EMS) model. The results indicated that the RN black hole is less thermodynamically favoured compared to the scalarized version. It should be noted that, as in the case of the Gauss-Bonnet coupling, the EMS models arise naturally in the low energy limit of string theory, while several black hole solutions have been found in this scenario \cite{Garfinkle:1990qj}, the most notable one being the Garfinkle, Horowitz, Strominger (GHS) black hole. In general the EMS models are governed by the following Lagrangian
\begin{equation}
    \mathcal{L} = \frac{R}{2} - \frac{1}{2}(\partial_{\mu}\phi)^2 - f(\phi)F_{\mu\nu}F^{\mu\nu} - V(\phi)~,
\end{equation}
where $f(\phi)$ is the coupling function (in string theory setups this is given by an exponential function for the dilaton) and  $V(\phi)$ is a scalar potential for the dilaton field. The dependence of the results on the form of the coupling functions has been addressed in \cite{Fernandes:2019rez}.
The stability of such black holes was assessed in \cite{Myung:2019oua,Blazquez-Salcedo:2020nhs}. Both electric and magnetic charges have been taken into account in \cite{Astefanesei:2019pfq} and scalarized black holes were obtained. The shadow casted by such black holes was analysed in \cite{Konoplya:2019goy} and the results unveiled that the scalarization always increases the radius of the shadow, regardless of the form of the coupling function. Axionic type couplings were considered in \cite{Fernandes:2019kmh}, while self interactions for the scalar field were introduced in \cite{Zou:2019bpt, Fernandes:2020gay} and the scalarization of charged black holes in the AdS case was discussed in \cite{Guo:2021zed}. Possible dynamical scalarization  in the RN-Melvin spacetime, which describes a charged black hole
permeated by a uniform magnetic field, was studied in \cite{Zhang:2022sgt} and a holographic scalarization of black holes with charged scalar fields was investigated in \cite{Guo:2020sdu}. Very recently, mixed spontaneous scalarization of EMS has also been explored in \cite{Belkhadria:2023ooc}.

As we have already discussed, the Gauss-Bonnet invariant arises from $\mathcal{O}(\mathcal{\alpha}')$ corrections to the bosonic sector of $\mathcal{N} =2$ supergravity, and the Gauss-Bonnet invariant is the $\mathcal{O}(\mathcal{\alpha}')$ $R^2$ correction to the Einstein action in the ten dimensional heterotic string theory \cite{Anninos:2008sj}. However, if one does not set the gauge field of the $U(1)$, that arises as a subgroup in the low energy effective theory of the heterotic string to zero, then higher order derivative gauge field terms coupled to the dilaton will arise \cite{Anninos:2008sj,Kats:2006xp}, such as:
\begin{equation}
    \sim h (F_{\mu\nu}F^{\mu\nu})^2~,\hspace{0.5cm} \sim h F^{~\sigma}_{\alpha}F^{~\alpha}_{\beta}F^{~\beta}_{\gamma}F^{~\gamma}_{\sigma}~, \label{motivation}
 \end{equation}
 where $h$ denotes the dilaton field. Such terms will arise as $\mathcal{O}(\mathcal{\alpha}')$ order corrections, just like the Gauss-Bonnet invariant. Moreover, by performing a non-diagonal reduction of the Gauss-Bonnet action, generating a gauge field in the lower-dimensional action one will also find terms like (\ref{motivation}) coupled to the dilaton field \cite{Charmousis:2012dw}. With the term non-diagonal reduction we refer to scenarios where the extra dimensions are not compactified and therefore the resulting metric cannot be brought to a block-diagonal form, where one block corresponds to the four dimensional space-time and another block that accounts for the extra dimensions. Consequently, these types of terms are well motivated from more fundamental string, as well as, higher dimensional gravity theories.

In this work we will consider the higher order derivative gauge field corrections that arise in the fundamental contexts we discussed previously. We will ignore higher order derivative terms for the scalar field, the Gauss-Bonnet invariant, as well as couplings of the derivatives of the scalar field with the Maxwell invariant and consider the Lagrangian
\begin{equation}
    \mathcal{L} = \frac{R}{2} - \frac{1}{2}(\partial_{\mu}\phi)^2 - \frac{1}{2}F_{\mu\nu}F^{\mu\nu} - f(\phi) F_{\mu\nu}F^{\mu\nu} + \alpha f(\phi) (F_{\mu\nu}F^{\mu\nu})^2~.
\end{equation}
The first terms (for $\alpha=0$) have already been considered in literature \cite{Fernandes:2019rez} and we will compare our findings with the existing results. One may arrive to such an action by field redefinitions as higher order derivative corrections to the GHS black hole. We did not include the scalar $F^{~\sigma}_{\alpha}F^{~\alpha}_{\beta}F^{~\beta}_{\gamma}F^{~\gamma}_{\sigma}$
since, as it is already pointed out in \cite{Natsuume:1994hd, Kats:2006xp}
this scalar provides a contribution similar to the $(F_{\mu\nu}F^{\mu\nu})^2$ scalar because of the fact that we will consider pure electric fields and hence we will not consider it here. We will then proceed in order to find the conditions under which the RN solution, which solves the theory for $\phi=0=f(0)$ is tachyonically unstable. We will see that for $\alpha>0$ the RN solution is unstable above a threshold of the parameter space and will admit a growing mode, once $f(\phi)>0$ and $\ddot{f}(\phi)>0$. We will use a quadratic coupling function to model our system and it is expected that more complicated functions that admit such expansion in the weak field limit will not affect (at least the onset of) the scalarization. Our results indicate that these black holes are thermodynamically preferred over the RN black hole and the branches do not end with an extremal black hole, but rather with a singularity as is evident from the behavior of the Kretschmann scalar. Moreover the energy conditions are respected for our spacetime.

This work is organized as follows: In Section II we provide a brief overview of the particular EMS model and we study the linear stability against scalar perturbations of the Reissner-Nordström black hole solution. In Section III we derive the non-trivial scalarized BH solutions and we examine their characteristics. In Section IV we discuss the thermodynamic properties of the scalarized BHs and in Section V we discuss their energy conditions. We summarize our findings and highlight potential future research attempts in Section VI.

\section{The EMS model with higher order derivative gauge field corrections}

In this section we will introduce our model, derive the field equations and perform an instability analysis in order to determine the conditions under which our theory develops a tachyonic instability.

\subsection{The set-up of the theory}
We consider the action of General Relativity, a scalar field with its kinetic term, which is non-minimally coupled to higher order derivative gauge field corrections, namely
\begin{equation}
    S= \frac{1}{8\pi}\int \, d^4x \, \sqrt{-g} \left[ \frac{R}{2}-\frac{1}{2} \nabla^\mu \phi \nabla_\mu \phi  -\frac{1}{2}\mathcal{P}- f(\phi) \left( \mathcal{P} -\alpha \mathcal{P}^2\right) \right]~, \label{action3}
\end{equation}
where
\begin{equation}
\mathcal{P} =  F_{\mu\nu}F^{\mu\nu} = -2(\mathbf{E}^2 -\mathbf{B}^2)~,
\end{equation}
and $F_{\mu\nu}=\partial_\mu A_\nu-\partial_\nu A_\mu$ is the Faraday tensor, $A_\mu$ is the gauge potential.
We will not consider dyons in this paper and as a result, the scalar $F^{~\sigma}_{\alpha}F^{~\alpha}_{\beta}F^{~\beta}_{\gamma}F^{~\gamma}_{\sigma}$ will be have similar contribution with the $\mathcal{P}^2$ scalar, and its absence will not affect qualitatively our results.  Consequently, the only quantity responsible for the non-linear electrodynamics modifications will be sourced by the scalar $\mathcal{P}^2$. We will use units in whose Newton's constant $G$ as well as the permeability of vacuum are normalized to unity.

By varying with respect to the dynamical fields $\phi,A^{\mu}$ and $g^{\mu\nu}$ respectively we can obtain the following equations of motion
\begin{eqnarray}
    &&\nabla_\mu \nabla^\mu \phi -\dot{f}(\phi) \left(\mathcal{P}-\alpha \mathcal{P}^2 \right)=0~, \label{KG} \\
    &&\nabla_\mu \left(2F^{\mu \nu}+f(\phi)F^{\mu\nu} -2\alpha f(\phi) \mathcal{P} F^{\mu\nu}\right) =0~,  \label{MAXWELLEQ} \\
    &&G_{\mu\nu} =  \mathcal{T}_{\mu\nu}^{SC}+\mathcal{T}_{\mu \nu}^{EM}+\mathcal{T}_{\mu\nu}^{INT}~,  \label{FIELDEQ}
\end{eqnarray}
where $\mathcal{T}_{\mu\nu}^{SC},~\mathcal{T}_{\mu \nu}^{EM},~\mathcal{T}_{\mu\nu}^{INT}$ are the energy momentum tensors of the scalar field,  the Maxwell invariant and the interaction term, and are given by
\begin{eqnarray}
    &&\mathcal{T}_{\mu\nu}^{SC} =  \nabla_\mu \phi \nabla_\nu \phi - \frac{1}{2} g_{\mu \nu} \nabla_\kappa \phi \nabla^\kappa \phi ~, \nonumber  \\
    &&\mathcal{T}_{\mu\nu}^{EM} = 2F_{\mu}^{~\kappa}F_{\nu \kappa} -\frac{1}{2}g_{\mu\nu} \mathcal{P}~, \nonumber \\
    &&\mathcal{T}_{\mu \nu}^{INT} =  f(\phi)\left(4F_{\mu}^{~\kappa}F_{\nu \kappa} -g_{\mu\nu} \mathcal{P}-8\alpha F_{\mu}^{~\kappa}F_{\nu \kappa} \mathcal{P} +\alpha g_{\mu\nu}\mathcal{P}^2 \right)~. \nonumber  \label{feqs}
\end{eqnarray}

The vacuum of our theory $\phi=0$ corresponds to the RN black hole upon considering $f(0)=\dot{f}(0)=0$. We wish to seek black hole solutions in the aforementioned theory, since it has been found that the much simpler scenario (when compared to a direct coupling of the scalar field to the Gauss-Bonnet invariant) of coupling a scalar field to electromagnetism might lead to scalarization of charged black holes \cite{Herdeiro:2018wub}. We study the  spontaneous scalarization process and we present the new scalarized black hole solutions and their properties. To do so we will numerically integrate the system of field equations with the appropriate boundary conditions.

\subsection{Spontaneous scalarization of RN black holes}

In order to scalarize the Reissner-Nordstrom black hole, we will first look at the behavior of a small perturbation around the vacuum of the scalar field theory
\begin{equation}
  ds^2 =-N(r) dt^2+\frac{1}{N(r)}dr^2+r^2d\theta^2+r^2\sin^2\theta d\varphi^2,~ N(r)\equiv 1-\frac{2M}{r}+\frac{Q^2}{r^2}~. \label{RNmetric}
\end{equation}
Hence we perform the perturbation $\phi \to 0 + \delta \phi$ and now the scalar equation of motion (\ref{KG}) becomes
\begin{equation} \left( \square - \mu_{\text{eff}}^2\right)\big|_{\phi=0}\delta \phi = 0~.\label{pertkg}
\end{equation}
where a term resembling an effective mass squared $\mu_{\text{eff}}^2$ for the perturbation reads as
\begin{equation}
  \mu_{\text{eff}}^2=  \ddot{f}(\phi)\left(\mathcal{P} - \alpha\mathcal{P}^2\right)\Big|_{\phi=0}~.
\end{equation}
If this effective mass squared is negative, the mass term becomes imaginary when squared. This implies that the perturbation oscillates with an exponentially growing or decaying amplitude. Of course, the perturbation which grows over time, indicates instability which is referred to as tachyonic instability. So the requirement of
the effective mass squared to be negative is a necessary condition for
spontaneous scalarization, but not sufficient, as in Minkowski spacetime, \cite{Doneva:2022ewd}. We observe that
$\mathcal{P}$ is always negative for pure electric fields (and therefore $\mathcal{P}^2>0$), $\alpha$ is related to the fine structure constant and is positive. As a result, to trigger a tachyonic instability, $\ddot{f}(\phi)\big|_{\phi=0}>0$ is required in order to scalarize the RN solution.
 Also according to the previous requirement that $f(0)=\dot{f}(0)=0$, we will consider the coupling function which is given by
\begin{equation}
    f(\phi)=\beta^2\phi^2,
\end{equation}
where $\beta$ is a dimensionless constant which shows the strength of the interaction.
Note that to obtain
a tachyonic instability it is necessary for the coupling function to be
quadratic when linearized around the scalar vacuum.\\
In order to determine the threshold of instability there are two possible ways, one is a dynamical (time-dependent) spherically symmetric perturbation and the other is a static (local) spherically symmetric perturbation, as in \cite{Myung:2018vug}. Because of the level of complexity we choose the second one. Performing a static (real) decomposition of the scalar field with the same symmetries of the RN background, namely $\delta\phi(r,\theta,\varphi)=u(r)Y_{lm}(\theta,\varphi)$, where $Y_{lm}(\theta,\varphi)$ are the spherical harmonic functions of degree $l$ and order $m$, the Eq.~(\ref{pertkg}) is reduced to the equation
\begin{equation}
   \left( r^2 N(r) u'(r)\right)' - \left( l(l+1)-\frac{2Q^2\left(2\alpha Q^2+r^4\right)f''(0)}{r^6}\right) u(r)=0~.
   \label{kgRN}
\end{equation}
We are interested in the spherically symmetric $l=0$ solutions, which are regular on and outside the horizon $r_H$ and vanish at infinity. When these unstable modes appear, the RN solution becomes unstable and new scalarized solutions with nontrivial scalar field bifurcate from it.
\begin{figure}[H]
    \centering
    \includegraphics[width=0.45\textwidth]{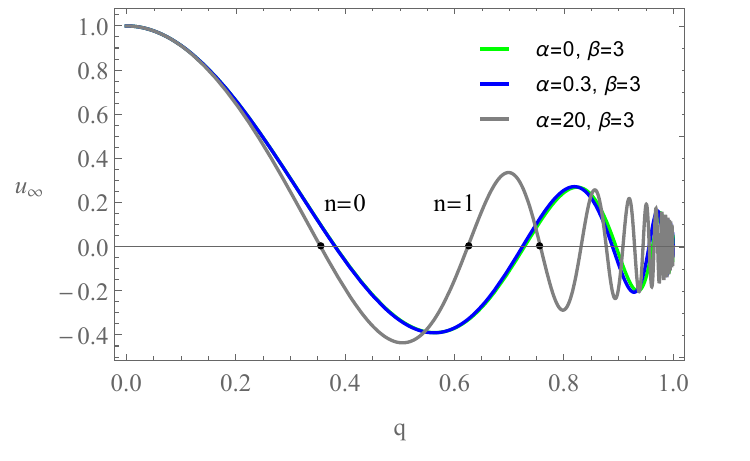}
    \includegraphics[width=0.45\textwidth]{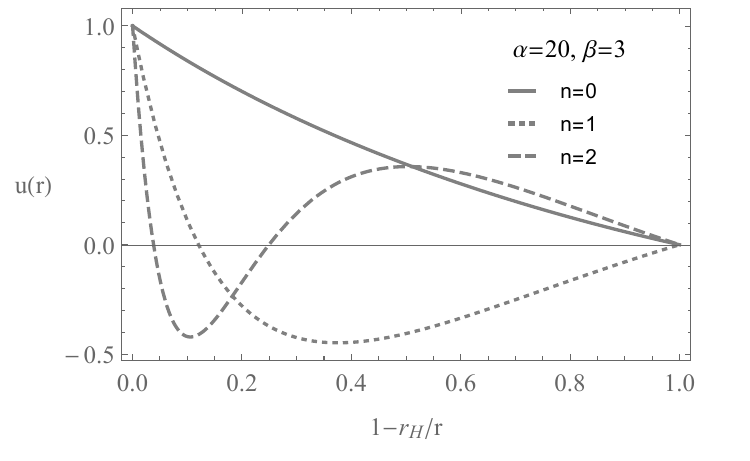}
    \caption{(Left) The value $u_{\infty}$ as a function of the charge to mass ratio $q$ of RN black hole. (Right) The radial profiles of perturbation $u(r)$ with different number of nodes.}
    \label{pertu}
\end{figure}
So in order to determine the regions of the parameter space where the RN solution is unstable, we solve numerically  Eq.~(\ref{kgRN}) and we study the value of the perturbation at infinity $u_{\infty}$, as in \cite{Astefanesei:2019pfq}. In Fig.~(\ref{pertu}) (left) we plot the $u_{\infty}$ as a function of the charge to mass ratio $q=\dfrac{Q}{M}$ of the RN black hole for different values of the coupling constants $\alpha,\beta$. The zeros of this function give us the unstable modes which characterized by a parameter $n=0,1,2,3,\dots$, which is associated with the number of nodes of $u(r)$, Fig.~(\ref{pertu}) (right). We explore the fundamental mode (zero mode, $n=0$), the first and the second mode ($n=1, n=2$ respectively) of the perturbations, \cite{Myung:2018vug}, \cite{Doneva:2010ke}. We call the existence value $q$, $q_{\text{exist}}$ as the smallest value of $q$ that onsets the instability of the RN solution.
\begin{figure}[H]
    \centering
    \includegraphics[width=0.45\textwidth]{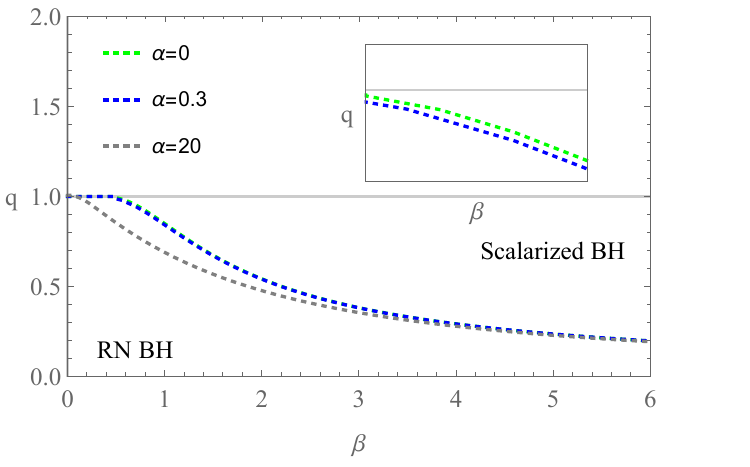}
    \includegraphics[width=0.4\textwidth]{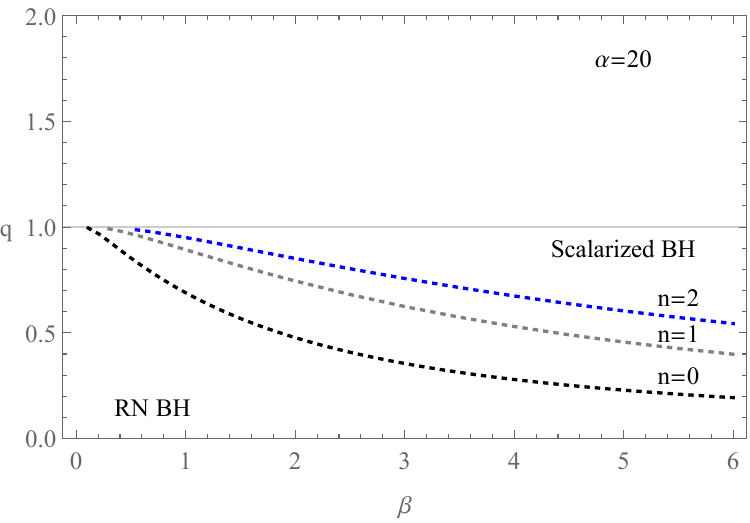}
    \caption{Lower threshold of domain of existence of scalarized BHs,  (left) for different values of the parameter $\alpha$ of the fundamental modes and (right) for three different modes with the same parameter $\alpha=20$}.
    \label{existenceplot}
\end{figure}
In Fig.~(\ref{existenceplot}) we demonstrate scalarized black hole (SBH) branches of solutions and we can see the lower threshold value $q_{\text{exist}}$ of the domain of existence of scalarized black hole solutions for different values of the constant $\alpha$. So the dotted lines separates the region where the RN black hole is stable (under dotted line) and the RN is unstable and scalarized solutions appear and bifurcate from the RN (above dotted line). As we can notice from Fig.~(\ref{existenceplot}) (left) sufficiently large values of constant $\alpha$ can increase the domain of existence of scalarized solutions. As it was expected from Fig.~(\ref{existenceplot}), \cite{Doneva:2010ke}, the fundamental mode ($n=0$) is described by smaller values of $q_{\text{exist}}$ and therefore the rest of the modes are less interesting.\\

\section{Spherically symmetric scalarized black hole solutions}

In this section we numerically solve the system of field equations. We highlight the primary findings, specifically focusing on the impact of the coupling of the scalar field to higher order derivative gauge field corrections on the domains of existence and the profiles of radial functions of scalarized black hole solutions.

\subsection{Asymptotic forms of the solutions at the horizon and at infinity}
In order to investigate scalarized charged black hole solutions we consider the following static and spherically symmetric ansatz for the metric
\begin{equation}
  ds^2 =-e^{-2\delta(r)}N(r) dt^2+\frac{dr^2}{N(r)}+r^2d\theta^2+r^2\sin^2\theta d\varphi^2,~ N(r)\equiv 1-\frac{2m(r)}{r}~, \label{metric ansatz}
\end{equation}
where $m(r)$ is the Misner-Sharp mass function and the gauge potential $A_{\mu}$
\begin{equation}
    A_{\mu}=\left( A(r),0,0,0\right) ~,\label{fourpotential}
\end{equation}
while the scalar field only depends on the radial coordinate, $\phi=\phi(r)$. \\
A linear combination of the Eq.~(\ref{KG}),(\ref{FIELDEQ}), using the integral of the Eq.~(\ref{MAXWELLEQ}) reads
\begin{eqnarray}
    &&  A'=-\frac{e^{\delta}r^2\left(1+2f(\phi)\right)}{2 \cdot 6^{1/3}\mathcal{C}}+\frac{e^{-3\delta}\mathcal{C}}{2 \cdot 6^{2/3}\alpha r^2 f(\phi)}~,\label{feqsVwithelectricalpotential1}\\
    && 2\delta'+r\phi'^2=0~,\\
    &&4 r A'^2 e^{2 \delta} f(\phi) \left(2 \alpha A'^2 e^{2 \delta }+1\right)-r \left(N''+N \left(-2 \delta ''+2 \delta '^2+\phi '^2\right)\right)+N'(r) \left(3 r \delta '-2\right)+2 \left(N \delta '+r A'^2 e^{2 \delta }\right)=0~,\\
    && N\phi''+\left(N'+\frac{N(2-r\delta')}{r}\right)\phi'+2e^{2\delta}\dot{f}(\phi)A'^2(1+2\alpha e^{2\delta}A'^2)=0~,\label{feqsV0withelectricalpotential}
\end{eqnarray}
where $\mathcal{C}=\mathcal{C}(r)$ reads as
\begin{equation}
\mathcal{C}=\left(\sqrt{6} \alpha^{3/2} r^4 e^{6 \delta } f(\phi)^{3/2} \sqrt{6 \left(9 \alpha Q^2+r^4\right) f(\phi )+4 r^4 f(\phi )^2 (2 f(\phi )+3)+r^4}+18 \alpha^2 Q r^4 e^{6 \delta} f(\phi )^2\right)^{1/3}~,
\end{equation}
where $Q$ is the integration constant which is interpreted as the electric charge. Note that the primes denote derivatives with respect to the radial coordinate.\\
To evaluate possible singular behaviors, it's noteworthy that the expressions for the Ricci and Kretschmann scalars, considering the line-element (\ref{metric ansatz}), are as follows:
\begin{eqnarray}
    && R=   \frac{N'}{r} (3r\delta'-4)+\frac{2}{r^2}\left(1+N\left(r^2\delta''-(1-r\delta')^2\right)\right)-N''~, \label{ricci}\\
    && K=   \frac{4}{r^4}(1-N)^2+\frac{2}{r^2}\left(N'^2+(N'-2N\delta')^2\right)+\left(N''-3\delta'N'+2N(\delta'^2-\delta'')\right)^2~.
    \label{krets}
\end{eqnarray}
We construct scalarized charged black hole solutions by integrating numerically the ordinary differential equations $(\ref{feqsVwithelectricalpotential1}-\ref{feqsV0withelectricalpotential})$ using a shooting method. At the black hole horizon $r=r_H$ the solutions are asymptotically flat and regular
\begin{eqnarray}
    &&m(r)=\frac{r_H}{2}+m'(r_H)(r-r_H)+\dots \nonumber\\
    &&\delta(r)=\delta(r_H)+\delta'(r_H)(r-r_H)+\dots \nonumber\\
     &&\phi(r)=\phi(r_H)+\phi'(r_H)(r-r_H)+\dots \nonumber\\
      &&A(r)=A(r_H)+A'(r_H)(r-r_H)+\dots
\end{eqnarray}
where
\begin{eqnarray}
    && m'(r_H)= \frac{1}{2}e^{2\delta(r_H)}r_H^2A'(r_H)^2 \left( 1+2f(\phi(r_H))\left(1+6\alpha e^{2 \delta(r_H)}A'(r_H)^2\right)\right)~, \\
    && \delta'(r_H)=-\frac{2e^{2\delta(r_H)}r_H^3 A'(r_H)^4 \dot{f}(\phi(r_H))\left(1+2\alpha e^{2\delta(r_H)}A'(r_H)^2\right)}{\left(-1+e^{2\delta(r_H)}r_H^2A'(r_H)^2 \left( 1+2f(\phi(r_H))\left(1+6\alpha e^{2 \delta(r_H)}A'(r_H)^2\right)\right)\right)^2}~, \\
    && \phi'(r_H)=\frac{2e^{2\delta(r_H)}r_H \dot{f}(\phi(r_H))A'(r_H)^2\left(1+2\alpha e^{2\delta(r_H)}A'(r_H)^2\right)}{-1+2e^{2\delta(r_H)}r_H^2A'(r_H)^2 \left( 1+f(\phi(r_H))\left(1+6\alpha e^{2 \delta(r_H)}A'(r_H)^2\right)\right)}~, \\
    && A'(r_H)=-\frac{e^{\delta(r_H)}r_H^2\left(1+2f(\phi(r_H))\right)}{2 \cdot 6^{1/3}\mathcal{C}(r_H)}+\frac{e^{-3\delta(r_H)}\mathcal{C}(r_H)}{2 \cdot 6^{2/3}\alpha r_H^2 f(\phi(r_H))} \label{Aprimerh}~.
\end{eqnarray}
The undetermined parameters $\delta(r_H),$ $\phi(r_H)$ and $A(r_H)$ are determined from the approximate behaviour of the solutions at large distances via shooting method. \\
At spatial infinity, the asymptotic solutions are
\begin{eqnarray}
    && m(r)=M-\frac{2Q^2+D^2}{4r}-\frac{M D^2}{4r^2} +\dots\\ \label{asymptotic0}
    && \delta(r)= \frac{D^2}{4r^2}+\frac{2M D^2}{3r^3}+\dots \label{asymptotic1} \\
    && \phi(r)=\frac{D}{r}+\frac{M D}{r^2}+\dots\\
    && A(r)=-\frac{Q}{r} +\dots \label{asymptotic}
\end{eqnarray}
where the parameters $M,Q,D$ denote, respectively, the ADM (Arnowitt, Deser, Misner) mass, the BH electric charge and the scalar charge at infinity.
The Ricci scalar $R$ (\ref{ricci}) approaches zero as $r$ approaches $r_H$, whereas the Kretschmann scalar $K$ (\ref{krets}) is expressed as follows:
\begin{eqnarray}
    K=&&\frac{12}{r_H^4} -\frac{24e^{\delta(r_H)}\left(1+2f(\phi(r_H))\right)A'(r_H)^2}{r_H^4}+ \frac{4e^{4\delta(r_H)}\left(5r_H^2+4f(\phi(r_H))\left(-14\alpha+5r_H^2+5r_H^2f(\phi(r_H))\right)\right)A'(r_H)^4}{r_H^2}\nonumber\\
    &&+352\alpha e^{6\delta(r_H)}f(\phi(r_H))\left(1+2f(\phi(r_H))\right)A'(r_H)^6+1600\alpha^2e^{8\delta(r_H)}f(\phi(r_H))^2A'(r_H)^8+\mathcal{O}(r-r_H)~.
\end{eqnarray}
where the $A'(r_H)$ is given by the Eq.~(\ref{Aprimerh}).

\subsection{Numerical results}
\begin{figure}[h]
    \centering
    \includegraphics[width=0.45\textwidth]{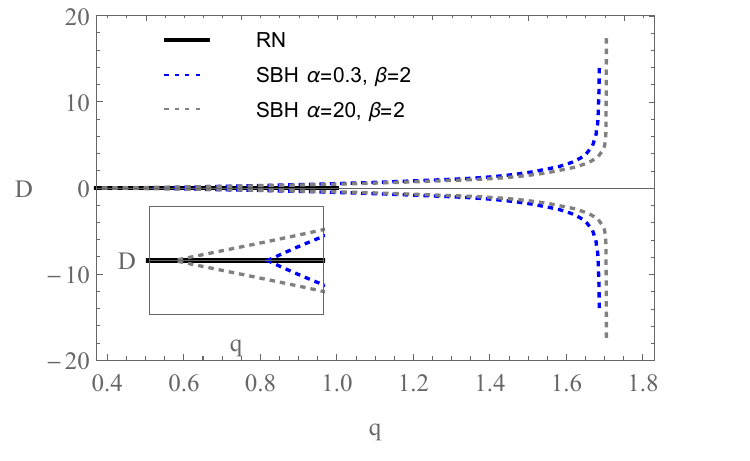}
    \includegraphics[width=0.4\textwidth]{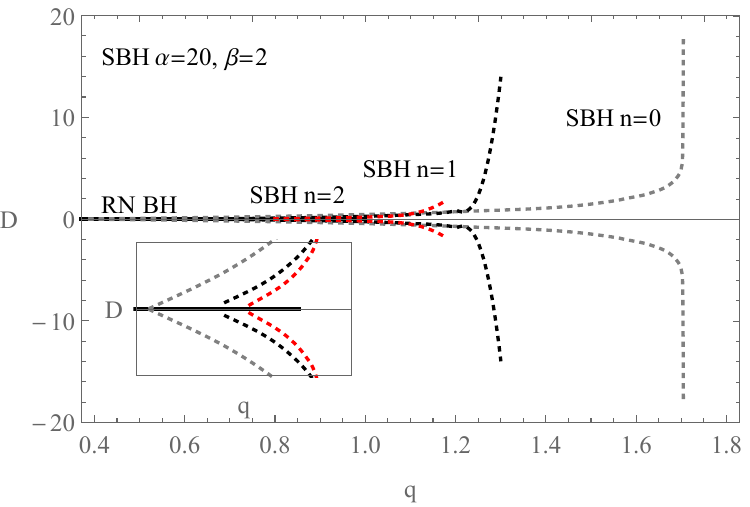}
    \caption{The scalar charge $D$ as a function of the charge to mass ratio $q$, (left) for different values of the parameters $\alpha, \beta$ of the fundamental modes and (right) for three different modes for the same parameters $\alpha, \beta$}.
    \label{plotD}
\end{figure}

\begin{figure}[h]
    \centering
    \includegraphics[width=0.45\textwidth]{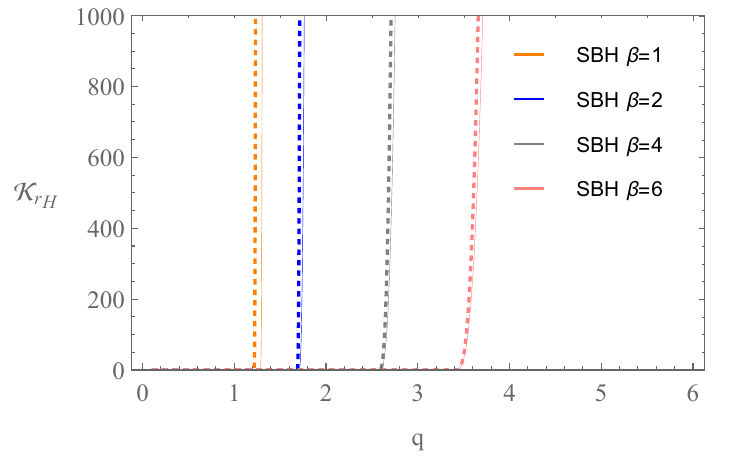}
    \includegraphics[width=0.45\textwidth]{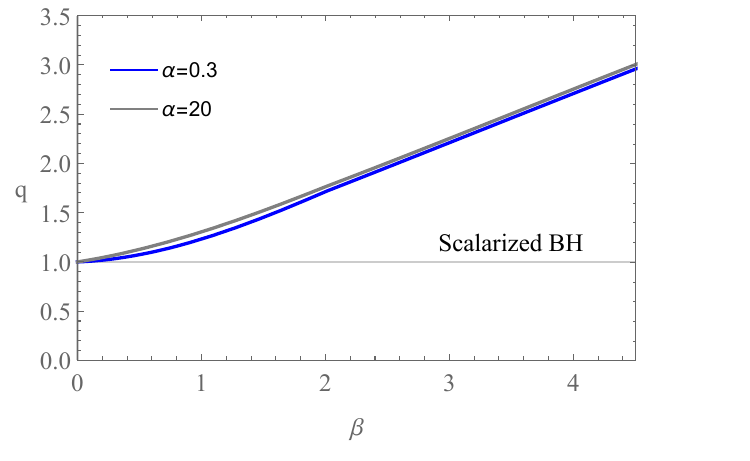}
    \caption{(Left) The Kretschmann scalar at the horizon  $\mathcal{K}_{r_H}$. The dotted lines describe solutions for $\alpha=0.3$ and the thin lines describe solutions for $\alpha=20$. As we move closer to the critical point, $\mathcal{K}_{r_H}$, diverges. (Right)  Upper threshold of domain of existence of scalarized BHs.}
    \label{plotDandkretch}
\end{figure}

\begin{figure}[h]
    \centering
    \includegraphics[width=0.45\textwidth]{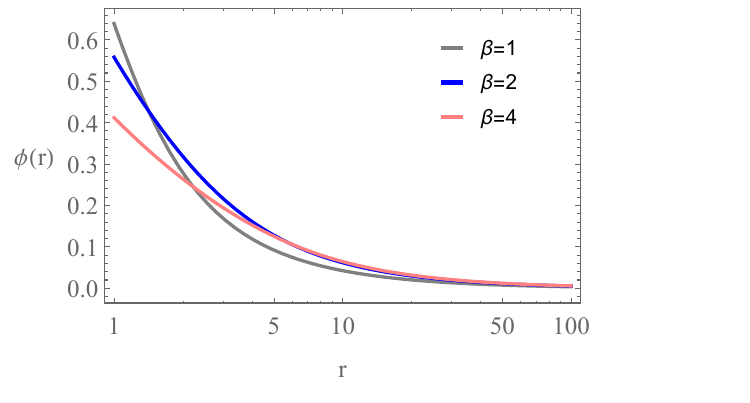}
    \includegraphics[width=0.43\textwidth]{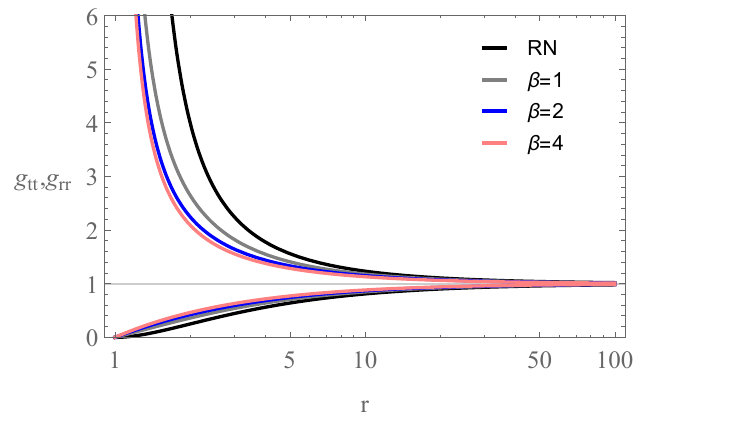}
    \includegraphics[width=0.43\textwidth]{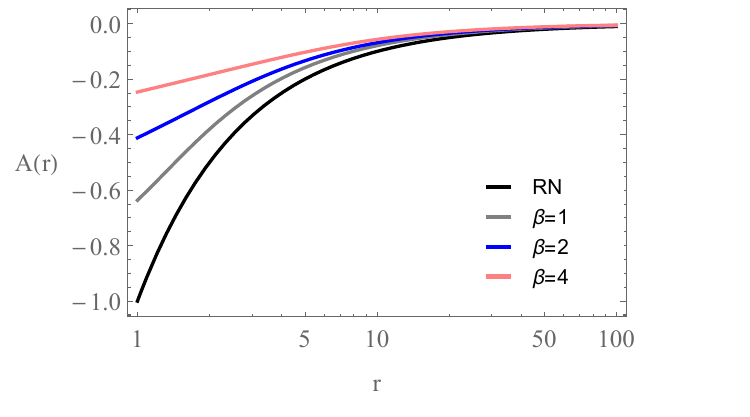}
    \caption{The scalar field $\phi(r)$, the metric $g_{tt},g_{rr}$ and the electric potential $A(r)$ as a function of the radial coordinate $r$. We set $\alpha=0.3,q=0.99$.}
    \label{plotbeta}
\end{figure}

The analysis of the linear stability has shown that the RN black hole has a tachyonic instability in a certain region of the parameter space, where we obtain numerically scalarized solutions bifurcating from it, as we can see in Fig.~(\ref{plotD}). Notice that each dot in the plot denotes a black hole solution. Each solution is found numerically, by solving the system of equations (\ref{feqsVwithelectricalpotential1})-(\ref{feqsV0withelectricalpotential}) with a shooting procedure in the Wolfram Mathematica software. The parameter that determines each solution is $\textbf{$r_H=1$}$ and there are three shooting parameters, namely, the value of the scalar field $\phi(r_H)$, the metric function $\delta(r_H)$ and the electric potential $A(r_H)$ at the horizon. The shooting method determines the aforementioned horizon quantities by the asymptotic solutions of $\phi, \delta, A$ at infinity, (\ref{asymptotic1})-(\ref{asymptotic}).
In Fig.~(\ref{plotD}) the thick black line denotes the trivial branch of Reissner–Nordström solution. Specifically, in Fig.~(\ref{plotD}) (left) the blue and gray dotted lines denote the nontrivial branches of scalarized black hole solutions for fundamental modes and in Fig.~(\ref{plotD}) (right) we demonstrate the first three nontrivial branches of the fundamental, the first and the second mode, respectively. As we can notice from the domain of existence of scalarized black holes (Fig.~(\ref{existenceplot})), the nontrivial branches bifurcate from the trivial branch and they can reach a charge to mass ratio $q$ greater than the unity. So scalarized black hole solutions can be overcharged, as they may have more electric charge than mass, while the black hole scalar charge increases to a critical value when the branch ends. The same happens for all the first three branches of nontrivial scalarized black hole solutions. Note also that the branch of the fundamental mode is bigger and tends with a greater charge to mass ratio than the other branches of the first and second modes. The scalar charge $D$ is, obviously, not independent from the black hole mass $M$, as the black hole charge $Q$, even if an explicit function that relates these quantities can not be found analytically, so the hair is of secondary kind. The endpoint of each branch exhibits a singularity, and numerical calculations indicate a divergence of the Kretschmann scalar at the horizon as we can notice in Fig.~(\ref{plotDandkretch}) (left). So we call the critical value of charge to mass ration $q$, $q_{\text{crit}}$ as the value which the Kretschmann scalar diverges. In Fig.~(\ref{plotDandkretch}) (right) we show the critical lines which  serve as upper bounds for the domain of existence of scalarized BHs.
The last one together with the Fig.~(\ref{existenceplot}) confirm that as the parameters $\alpha, \beta$ are increasing the domain of existence of scalarized black holes also is increasing. In Table.~(\ref{table}) we show the existence and critical values of charge to mass ratio, $q_{\text{exist}}, q_{\text{crit}}$ respectively, for some branches of scalarized solutions.
\begin{table}[H]
    \centering
    \begin{tabular} { | m{1cm} | m{1cm}| m{2cm} |  m{2cm}| }
    \hline
        $\alpha$ &   $\beta$   & $q_{\text{exist}}$ & $q_{\text{crit}}$ \\
        \hline \hline
        $0.3$    &      $2$    &  $0.54079$  & $1.68719$  \\
        \hline
        $20$     &      $2$    &  $0.47737$  & $1.70369$  \\
        \hline
        $0.3$    &      $4$    &  $0.29235$  & $2.58449$  \\
        \hline
        $20$     &      $4$    &  $0.27943$  & $2.58737$  \\
        \hline
    \end{tabular}
    \caption{Threshold values of charge to mass ratio, $q_{\text{exist}}, q_{\text{crit}}$, for different branches.}
    \label{table}
\end{table}
We can notice from the threshold values, $q_{\text{exist}}, q_{\text{crit}}$, that the effect of the coupling constant $\beta$ is more significant than the effect of the constant $\alpha$, in the sense that small value changes of $\beta$ will result in configurations with bigger deviations when compared to RN. In Fig.~(\ref{plotbeta}) we show such configurations for $\alpha=0.3,~q=0.99$ and $\beta=1,2,4$.
As we can notice the scalar field configurations are characterized by no appearance of zeros. All the configurations deviate from each one as the coupling constant $\beta$ is increasing not only qualitatively but also quantitatively. The value of the scalar field at the horizon is decreasing while the value at infinity approaches its asymptotic value with a slower rate, as $\beta$ is increasing. The components of the metric $g_{tt}, g_{rr}$ of scalarized solutions, as well as the electric potential $A(r)$, demonstrate significant deviation from the  Reissner–Nordström one, Fig.~(\ref{plotbeta}) (right), (down).
\begin{figure}[h]
    \centering
    \includegraphics[width=0.45\textwidth]{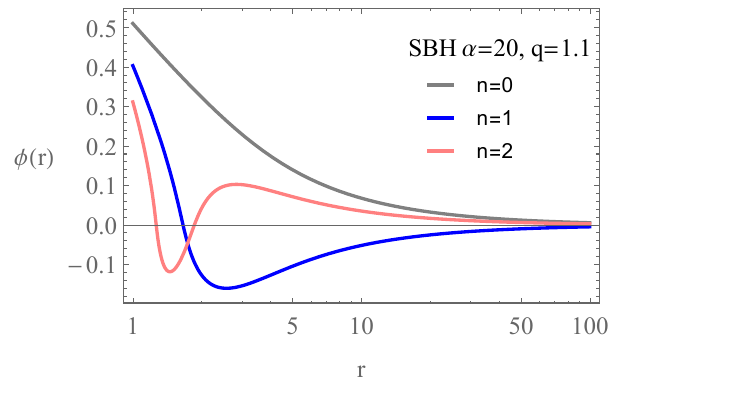}
    \caption{The scalar field $\phi(r)$ as a function of the radial coordinate $r$ for the first three modes. We set $\alpha=20,q=1.1$.}
    \label{plotscalarmodes}
\end{figure}
In Fig.~(\ref{plotscalarmodes}) we depict three scalar field configurations $\phi(r)$ for the first three modes, where we can notice the zeros of each mode. The fundamental mode does not develop any root, while the first and the second modes do develop one and two roots respectively.

\section{Thermodynamics and Smarr Relation}
Let us now discuss the thermodynamics of the solution obtained. We are dealing with a stationary, asymptotically flat spacetime, which therefore admits an asymptotically timelike vector field $K^{\mu} = (1,0,0,0)$, which satisfies the Killing equation $\nabla_{\mu}K_{\nu}+\nabla_{\nu}K_{\mu}=0$. As a result we can define the conserved mass of the black hole as \cite{Poisson:2009pwt}
\begin{equation}
    \mathcal{M} = -\frac{1}{8\pi}\lim_{r\to\infty}\int_{\infty} d S_{\alpha\beta}\nabla^{\alpha}K^\beta~, \label{conservedmass}
\end{equation}
where $dS_{\alpha\beta}  = -2t_{[\alpha}r_{\beta]}\sqrt{\sigma}d\theta d\varphi$ is the surface element with $\sqrt{\sigma}$ being the induced metric on the $t=r=\text{const}$ surface: $\sqrt{\sigma} = r^2 \sin\theta$. Here $t_\mu$ is a time-like covariant vector field, normalized to unity $t_{\mu} =( -\sqrt{e^{-2\delta}N},0,0,0)$ and $r_\mu$ is a space-like covariant vector field normalized to satisfy $r_\mu = (0,N(r)^{-1/2},0,0)$. Expanding the term $d S_{\alpha\beta}\nabla^{\alpha}K^\beta$ we have
\begin{equation}
    d S_{\alpha\beta}\nabla^{\alpha}K^\beta =-2t_{[\alpha}r_{\beta]}\nabla^{\alpha}K^\beta \sqrt{\sigma}d\theta d\varphi= (-t_\alpha r_\beta \nabla^{\alpha}K^\beta + t_{\beta}r_{\alpha}\nabla^{\alpha}K^\beta)\sqrt{\sigma}d\theta d\varphi= - 2t_{\alpha}r_{\beta}\nabla^{\alpha}K^\beta \sqrt{\sigma}d\theta d\varphi~.
\end{equation}
Evaluating the above relation for our line element we have that
\begin{equation}
    2t_{\alpha}r_{\beta}\nabla^{\alpha}K^\beta = -2t_tr_r\Gamma_{tt}^rK^t \sim - 2 M r^2~,
\end{equation}
where we have used the asymptotic form of the solution given in equations (\ref{asymptotic0})- (\ref{asymptotic}) and kept only the highest order term, since the integral is evaluated at a $2-$sphere at infinity. Finally evaluating the integral (\ref{conservedmass}) we obtain
\begin{equation}
    \mathcal{M} = -\frac{1}{8\pi}(-8\pi M) = M~,
\end{equation}
which ensures that indeed $M$ is the ADM mass as measured by a far-away observer.

Now, since the Killing equation is antisymmetric, it satisfies the following idenity
\begin{equation}
    \oint_{\partial \Sigma} \nabla^{\alpha}K^{\beta} dS_{\alpha\beta} = 2 \int_{\Sigma}\nabla_\beta\nabla^{\alpha}K^{\beta}d\Sigma_{\alpha}~, \label{antisymmetry}
\end{equation}
which might be re-written as
\begin{equation}
    \oint_{\partial \Sigma} \nabla^{\alpha}K^{\beta} dS_{\alpha\beta} = 2 \int_{\Sigma} R^{\alpha}_{~\beta}K^{\beta}d\Sigma_{\alpha}~, \label{antiricci}
\end{equation}
if one uses the antisymmetric nature of the Killing equation, as well as, the equation $\square K^a = -R^{a}_{~b}K^b$.
The left hand side of (\ref{antisymmetry}) contains two contributions from the cross-section defined by $t=r=\text{const}$ one at the event horizon of the black hole and another one at infinity. As a result we can break this term into two pieces
\begin{equation}
    \oint_{\partial \Sigma} \nabla^{\alpha}K^{\beta} dS_{\alpha\beta} = \oint_{H} \nabla^{\alpha}K^{\beta} dS_{\alpha\beta}+
    \oint_{\infty} \nabla^{\alpha}K^{\beta} dS_{\alpha\beta}~,
\end{equation}
and we have already calculated the term at infinity, which will give $-8\pi M$. Evaluating the integral at the horizon we have
\begin{equation}
    \oint_{H} \nabla^{\alpha}K^{\beta} dS_{\alpha\beta} = 4\pi r^2 e^{-\delta}N'\Big|_{r_H}~.
\end{equation}
As a result one may now write
\begin{equation}
    -8\pi M + 4\pi r^2 e^{-\delta}N'\Big|_{r_H} = 2 \int_{\Sigma} R^{\alpha}_{~\beta}K^{\beta}d\Sigma_{\alpha} \to M = \frac{1}{2} r^2 e^{-\delta}N'\Big|_{r_H} -\frac{1}{4\pi}\int_{\Sigma} R^{\alpha}_{~\beta}K^{\beta}d\Sigma_{\alpha}~. \label{smarr1}
\end{equation}
The area of the event horizon of the black hole is given by \cite{Poisson:2009pwt}
\begin{equation}
    \mathcal{A}(r_H) = \int_{0}^{2\pi}d\varphi\int_{0}^{\pi}r_H^2\sin\theta =4\pi r_H^2~.
\end{equation}
The temperature of the black hole at the event horizon is $T_H = N'e^{-\delta}/4\pi\Big|_{r_H}$ \cite{Poisson:2009pwt}. Now we can rewrite (\ref{smarr1}) as
\begin{equation}
    M = \frac{1}{2}\mathcal{A}T - \frac{1}{4\pi}\int_{\Sigma} R^{\alpha}_{~\beta}K^{\beta}d\Sigma_{\alpha}~.
\end{equation}
Moreover, the $t=\text{const}$ hypersurface element reads
\begin{equation}
    d\Sigma_{\alpha} = - t_{\alpha}\sqrt{h}~,
\end{equation}
where $h=r^2 \sin \theta \sqrt{1/N}$ is the induced metric on the spacelike hypersurface.
Now, by using Einstein's equation we may rewrite the above equation as
\begin{equation}
    M = \frac{1}{2}\mathcal{A}T + \frac{1}{4\pi}\int_{r_H}^{\infty}dr\int_{0}^{\pi}\sin\theta d\theta\int_{0}^{2\pi}d\varphi\left\{-\sqrt{e^{-2\delta}N}r^2\sqrt{\frac{1}{N}}\left(-e^{2 \delta } \left(A'\right)^2 \left(f(\phi ) \left(4 \alpha  e^{2 \delta } \left(A'\right)^2+2\right)+1\right)\right)\right\}~, \label{smarr2}
\end{equation}
where we have used the trace of the energy-momentum tensor
\begin{equation}
    \mathcal{T} = -16 \alpha  e^{4 \delta } \left(A'\right)^4 f(\phi )-r \left(\phi '\right)^2~.
\end{equation}
Notice here the absence of any $A'(r)^2$ term because of the fact that the Maxwell's theory is traceless in four dimensions.  Now (\ref{smarr2}) reads
\begin{equation}
     M = \frac{1}{2}\mathcal{A}T + \int_{r_H}^{\infty}dr\left(e^{\delta } r^2 \left(A'\right)^2 (2 f(\phi )+1) +4 \alpha  e^{3 \delta } r^2 \left(A'\right)^4 f(\phi ) \right)~,
\end{equation}
and this is the Smarr relation that our solution satisfies. For the free scalar field theory where $\phi=f(\phi)=\delta=0$ one can see that
\begin{equation}
    M = \frac{1}{2}\mathcal{A}T + \Phi_{RN} Q~,
\end{equation}
where $\Phi_{RN} = Q/r_H$ is the electrostatic potential of the RN black hole, and hence one obtains the usual Smarr formula.

The charge of the scalar field might also be computed by using the relation of the dilaton charge, mostly used in string theory \cite{Kanti:1995vq, Garfinkle:1990qj}
\begin{equation}
    D= -\frac{1}{4\pi}\int d^2\Sigma^{\mu}\nabla_{\mu}\phi~,
\end{equation}
where the integral is evaluated over a two-sphere with infinite radius and $-1/4\pi$ is a normalization constant. It might not be clear from this expression, however, the scalar field dresses the black hole with a secondary scalar hair, since the scalar charge is not independent from the mass of the black hole, or the electric charge, as we mentioned above and as can be clearly seen in Fig.~(\ref{plotD}).

Moreover, as is well known, the entropy will be related to the area of the black hole solution \cite{Bekenstein:1973ur}. It has also been proven that, the entropy will be associated to the gravitational theory under consideration through Wald's formula \cite{Wald:1993nt}.  In this work we considered the framework of General Relativity to describe gravitation and consequently the entropy will be given by
\begin{equation}
    \mathcal{S} = \frac{\mathcal{A}(r_H)}{4}~,
\end{equation}
since we have set Newton's constant to unity. As a result, examining the area of the black hole is the same as examining the entropy.

\subsection{The thermodynamic quantities}
\begin{figure}[H]
    \centering
    \includegraphics[width=0.45\textwidth]{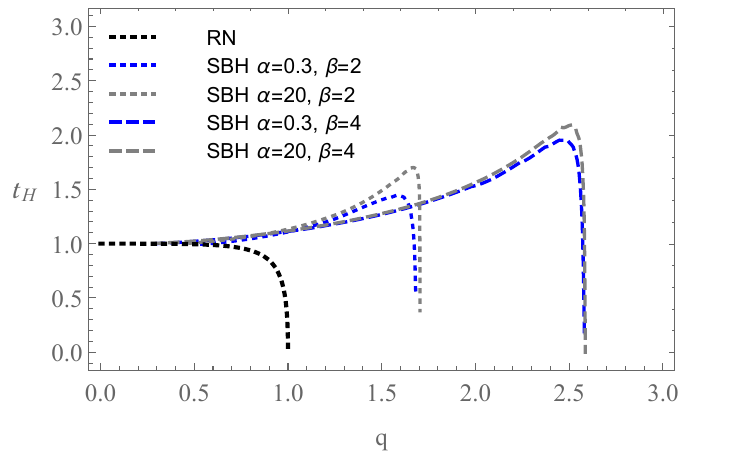}
    \includegraphics[width=0.45\textwidth]{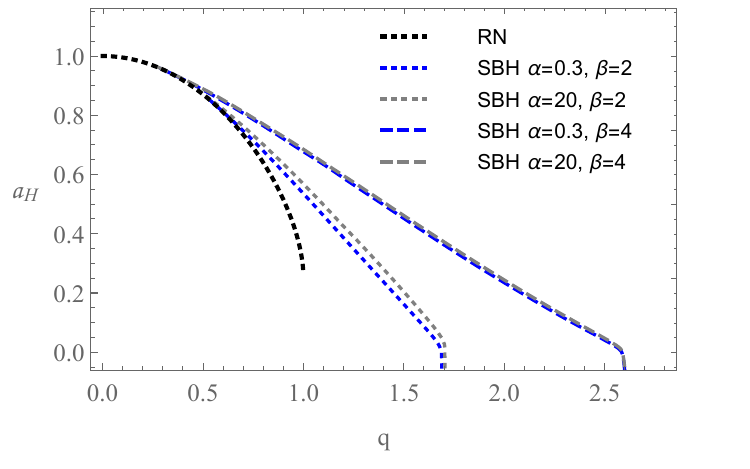}
    \caption{(Left) Reduced temperature $t_H$ as a function of the charge to mass ratio $q$. (Right) Reduced area $a_H$ as a function of the charge to mass ratio $q$.}
    \label{plottemparea}
\end{figure}

\begin{figure}[H]
    \centering
    \includegraphics[width=0.4\textwidth]{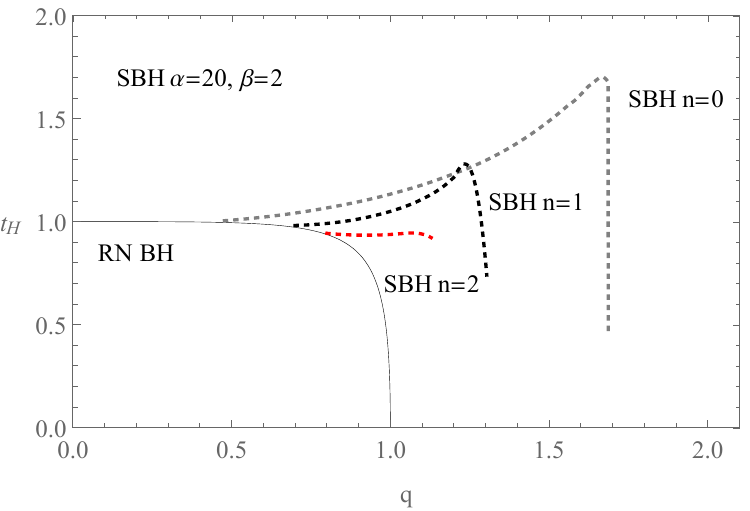}
    \includegraphics[width=0.4\textwidth]{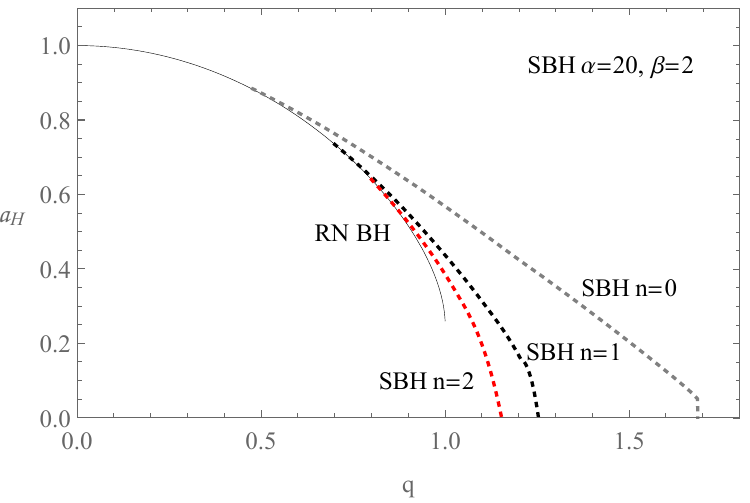}
    \caption{(Left) Reduced temperature $t_H$ as a function of the charge to mass ratio $q$ for the first three nontrivial branches. (Right) Reduced area $a_H$ as a function of the charge to mass ratio $q$. for the first three nontrivial branches.}
    \label{plottemparean0n1n2}
\end{figure}

We introduce the dimensionless standard reduced quantities
\begin{equation}
    a_H \equiv \frac{\mathcal{A}_H}{16 \pi M^2} \, , \, t_H\equiv 8 \pi T_H M~.
\end{equation}
In Fig.~(\ref{plottemparea}) we plot the reduced temperature and the area of RN black hole solution as well as of some scalarized branches of solutions.
As we discussed above, we can notice in Fig.~(\ref{plottemparea}), that for a given set of constants $\alpha$ and $\beta$, nontrivial scalarized black holes emerge through bifurcation from the corresponding Reissner-Nordström black hole with a specific charge to mass ratio $q_{\text{exist}}$. The branches of solutions have a finite range and end up at a critical configuration with a different ratio $q_{\text{crit}}$. The resulting solution features a singular horizon, evidenced by the evaluation of the Kretschmann scalar (Fig.~(\ref{plotDandkretch}) (left)). As the critical solution is approached, the horizon area tends to zero and the temperature remains finite and is decreasing as long as the coupling is getting stronger as well. It is essential to mention that there are BHs which are hot as indicated by the peaks in the reduced temperature plot. In the parameter space region where both scalarized and RN black holes coexist for the same charge $q$, it is consistently observed that scalarized solutions are entropically favored over RN black holes, as evident in Fig.~(\ref{plottemparea}) (right).
In Fig.~(\ref{plottemparean0n1n2}) we can notice the same behavior of the reduced temperature and area of all of the first three nontrivial branches. The fundamental mode exhibits higher entropy in comparison to the rest of the modes, so the non-fundamental modes are not thermodynamically preferred. 

\section{Energy Conditions}

In this section we will discuss the nature of the energy momentum tensor threading the black hole spacetime, by analyzing the energy conditions \cite{Kontou:2020bta}. By considering the proper reference frame where an observer will remain at rest for constant $r,\theta,\varphi$ \cite{Morris:1988cz},  we may identify the energy density and the principal pressures are follows
\begin{eqnarray}
    && \rho \equiv -\mathcal{T}^{t}_{~t} = e^{2 \delta } \left(A'\right)^2 \left(2 f(\phi ) \left(6 \alpha  e^{2 \delta } \left(A'\right)^2+1\right)+1\right)+\frac{1}{2} N \left(\phi
   '\right)^2~,\\
   &&p_r \equiv \mathcal{T}^{r}_{~r} = \frac{1}{2} N \left(\phi '\right)^2-e^{2 \delta } \left(A'\right)^2 \left(2 f(\phi ) \left(6 \alpha  e^{2 \delta }\left(A'\right)^2+1\right)+1\right)~,\\
   &&p_{\theta} = p_{\varphi} \equiv \mathcal{T}^{\theta}_{~\theta} = e^{2 \delta } \left(A'\right)^2 \left(f(\phi ) \left(4 \alpha  e^{2 \delta } \left(A'\right)^2+2\right)+1\right)-\frac{1}{2} N \left(\phi
   '\right)^2~.
\end{eqnarray}

\begin{figure}
    \centering
    \includegraphics[width=0.48\textwidth]{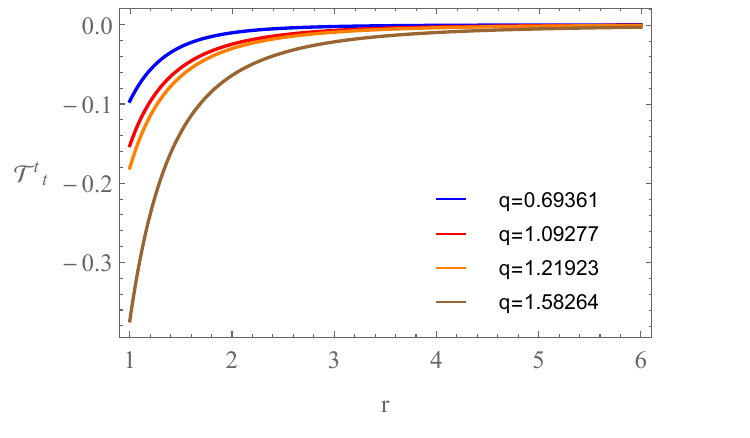}
    \includegraphics[width=0.48\textwidth]{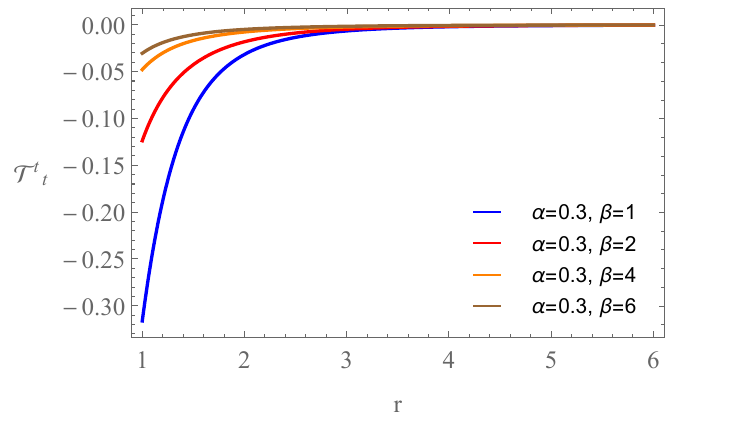}
    \includegraphics[width=0.48\textwidth]{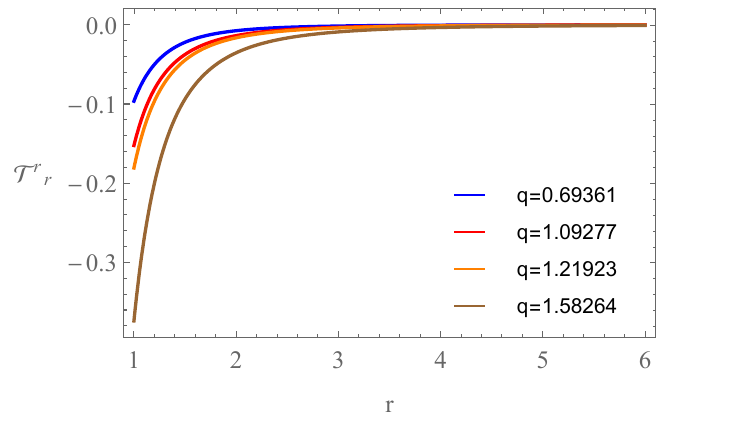}
    \includegraphics[width=0.48\textwidth]{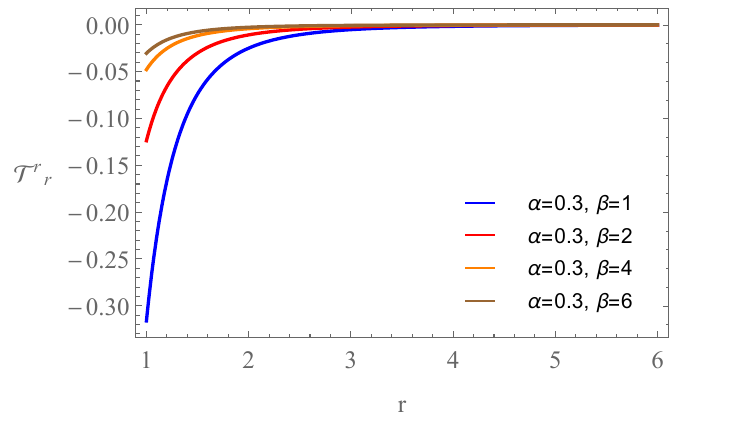}
    \includegraphics[width=0.48\textwidth]{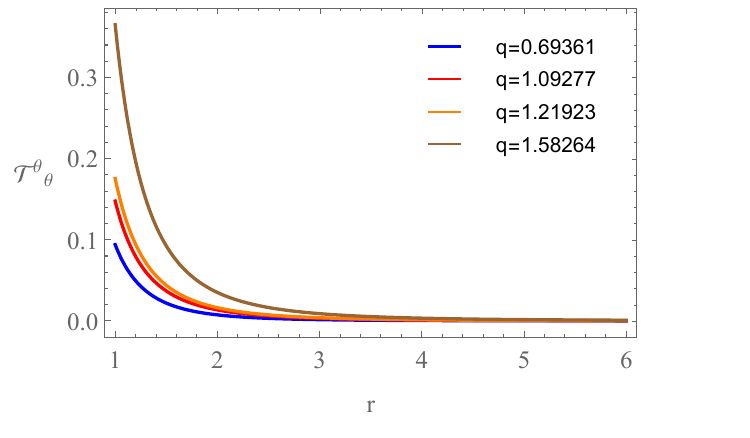}
    \includegraphics[width=0.48\textwidth]{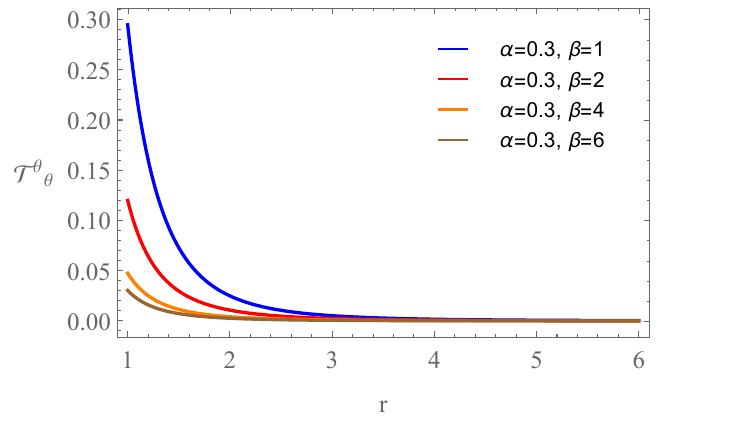}
    \caption{The components of the energy-momentum tensor for scalarized BHs for different scenarios.(Left) We set $\alpha=0.3, \beta=2$ (Right) We set $q=0.99$.}
    \label{plotemt}
\end{figure}

Without referring to the exact form of the solutions, the energy density of the black hole spacetime is always positive \textit{by construction} in the exterior region of the black hole $r>r_H$ where $N>0$, since in order to have scalarized solutions we assumed that $f(\phi)>0$ for a positive $\alpha$. As a result, the Weak Energy Condition which implies the non-negativity of the energy density, is respected. Moreover, the Null Energy Condition (NEC) states that the sum of the energy density with the radial pressure is non-negative. For our scenario we have
\begin{equation}
    \rho+p_r = N\phi'^2~,
\end{equation}
which is clearly positive in the causal region of spacetime and the NEC again holds \textit{by construction} since we used a regular scalar field in order to construct hairy black hole solutions and not a phantom one (with a negative kinetic energy term in the Lagrangian). The Strong Energy Condition (SEC) states that the sum of the energy density and the principal pressures is non-negative which for our case reads
\begin{equation}
    \rho+p_r +p_\theta+p_\varphi=2 e^{2 \delta } \left(A'\right)^2 \left(f(\phi ) \left(4 \alpha  e^{2 \delta } \left(A'\right)^2+2\right)+1\right)~,
\end{equation}
which is also non negative for our system. Hence the WEC, NEC and SEC are all satisfied in the causal region of spacetime for our solution, since the pressure of the matter threading the black hole spacetime is tangential dominated \cite{Dorlis:2023qug}. In Fig. ~ (\ref{plotemt}) we plot the components of the energy-momentum tensor of our theory. It is clear that all components are finite at the event horizon of the black hole while at infinity tend to zero, in accordance with asymptotic flatness.
 In Fig.~(\ref{plotemt}) (left) we set $\alpha=0.3$ and $\beta=2$ so we plot the components $\mathcal{T}^{\mu}_{~\mu}$ for some scalarized BH solutions with different charge to mass ratio $q$. We can notice that as the BHs are getting overcharged, the magnitude of all components is increasing at the horizon, while it reaches its asymptotic value at a slower rate. In the right column we set $\alpha=0.3, q=0.99$ while we increase the strength of the interaction of the scalar field with the electromagnetism. As the coupling constant $\beta$ is increasing the magnitude of all the components is decreasing.

\section{Conclusions}

In this work we considered the EMS model with higher derivative gauge field corrections, a scenario that arises in string theory setups, as well as dimensionally reduced Lovelock theories. We investigated the conditions under which the background solution of our theory (the RN black hole) develops a tachyonic instability, indicating in this way the spontaneous dressing of the RN black hole with the scalar field of the theory. Then we solved numerically the full field equations and found that we have scalarized black hole  solutions that carry a non-trivial scalar field. The branches of our black hole solutions end with a curvature singularity and not with an extremal black hole, which is in agreement with \cite{Doneva:2018rou}. We investigated the thermodynamics of our system, derived the Smarr relation of our black hole spacetime  and defined the mass and the scalar charge of our solution through hypersurface integrals. By examining the temperature of the black hole, we found that there exists a critical value of the electric charge to mass ratio for which the black holes are hot. The area of the scalarized black hole solutions is bigger when compared to the area of the RN black hole, as well as, to the area of the EMS scalarized black holes without the higher derivative gauge field corrections. This result indicates that our solutions are thermodynamically preferred  when compared to the existing literature.

One can now build upon our results by introducing the terms that we omitted, such as the Gauss-Bonnet term, the coupling of derivatives of the scalar field with the Maxwell invariant etc. It would be of interest to consider dyons. In this case, the addition of the scalar $F^{~\sigma}_{\alpha}F^{~\alpha}_{\beta}F^{~\beta}_{\gamma}F^{~\gamma}_{\sigma}$ will not lead to the same contributions as the $(F_{\mu\nu}F^{\mu\nu})^2$ term we considered here, so one could compare our findings with the dyonic case. Following  \cite{Blazquez-Salcedo:2020nhs}, the stability of the scalarized black hole solutions may be of interest to be addressed.

\section{Acknowledgments}
The research project was supported by the Hellenic Foundation for Research and Innovation (H.F.R.I.) under the “3rd Call for H.F.R.I. Research Projects to support Post-Doctoral Researchers” (Project Number: 7212).
We would like to thank Athanasios Bakopoulos, Cristian Erices, Carlos A.R. Herdeiro, Alexandre M. Pombo and Stoytcho Yazadjiev for valuable discussions. We also thank an anonymous referee for valuable comments and suggestions that improved the quality of our paper.

\end{document}